\documentclass[pdftex,12pt]{article}
\pdfoutput=1
\usepackage{natbib}
\usepackage[pdftex]{graphicx}        
\usepackage{color}           
\usepackage{url}             
\usepackage[breaklinks=true]{hyperref}

\usepackage{mbenotes}

\usepackage{txfonts}
\usepackage[font={small}]{caption}

\title{\vspace{-3cm}Oscillations above sunspots from the temperature minimum to the corona}

\author{N.I.~Kobanov, A.A.~Chelpanov, and D.Y.~Kolobov\\
    \small{Institute of Solar-Terrestrial Physics} \\
    \small {of Siberian Branch of Russian Academy of Sciences, Irkutsk, Russia} \\
    \small {email: \url{kobanov@iszf.irk.ru}}
    }
\date{\small{[
{This article was accepted for publication in \textit{Astronomy\&Astrophysics}}]}}

\begin{document}
\maketitle

\begin{abstract}
  \par \textbf{Context.}{ An analysis of the oscillations above sunspots was carried out using
simultaneous ground-based and Solar Dynamics Observatory (SDO) observations
(Si\,\textsc{i}~10827\,\AA,
He\,\textsc{i}~10830\,\AA, Fe\,\textsc{i}~6173\,\AA, 1700\,\AA,
He\,\textsc{ii}~304\,\AA, Fe\,\textsc{ix}~171\,\AA).}
  \par \textbf{Aims.}{ Investigation of the spatial distribution of oscillation power in the
frequency range 1--8\,mHz for the different height levels of the solar atmosphere.
Measuring the time lags between the oscillations at the different layers.}
  \par \textbf{Methods.}{ We used frequency filtration of the intensity and Doppler velocity 
  variations with Morlet wavelet
to trace the wave propagation from the photosphere to the chromosphere and the
corona.}
  \par \textbf{Results.}{ The 15 min oscillations are concentrated near the outer penumbra in the upper
photosphere (1700\,\AA), forming a ring, that expands in the transition zone.
These oscillations propagate upward and reach the corona level, where their
spatial distribution resembles a fan structure. The spatial distribution of
the 5 min oscillation power looks like a circle-shape structure matching the sunspot
umbra border at the photospheric level. The circle expands at the higher levels
(He\,\textsc{ii}~304\,\AA\ and Fe\,\textsc{ix}~171\,\AA). This indicates that the
low-frequency oscillations propagate along the inclined magnetic tubes in the
spot. We found that the inclination of the tubes reaches 50--60 degrees in the upper
chromosphere and the transition zone.
\par The main oscillation power in the 5--8\,mHz range concentrates within the
umbra boundaries at all the levels. The highest frequency oscillations (8\,mHz)
are located in the peculiar points inside the umbra. These points  probably coincide with
umbral dots.
\par We deduced the propagation velocities to be 28$\pm$15
$\mathrm{km\,s^{-1}}$, 26$\pm$15 $\mathrm{km\,s^{-1}}$, and 55$\pm$10
$\mathrm{km\,s^{-1}}$ for the Si\,\textsc{i}~10827\,\AA--He\,\textsc{i}~10830\,\AA,
1700\,\AA--He\,\textsc{ii}~304\,\AA, and
He\,\textsc{ii}~304\,\AA--Fe\,\textsc{ix}~171\,\AA\ height levels, respectively.
}

\end{abstract}

\section{Introduction}
 \par Sunspot oscillations in the lower solar atmosphere have been observed and discussed for many decades
\citep{beckers1969uf,zirin1972,giovanelli1972,1990A&A...237..243B,
1985A&A...143..201Z,lites1992a, tsiropoula2000}.
The problem has turned out to be complicated \citep{bogdan2006a,2008sust.book.....T}.
Despite some success in this field, we
are still far from a comprehensive understanding of the wave processes in and
above sunspots.

\par Chromospheric waves in sunspots are readily detected in the H$\alpha$
and He\,\textsc{i} 10830\,\AA\ lines. It was found that different oscillatory
modes coexist and the dominant frequency changes from 6--7 mHz in the umbra
to 1.5--2 mHz in the outer penumbra. One hypothesis to explain the observed wave behavior is
the common source located in subphotospheric layers. The scenario is that the
waves originate in the deep layers of the solar atmosphere and propagate upward
along the magnetic field lines \citep{roupe2003,kob2006,kob2011,bloomfield2007}.
In this case, one expects a positive time delay for the waves observed at the upper
and lower levels. Detecting such a delay is a complex task, and the results from
different authors are not always consistent with one another
\citep{kob2011,centeno2009}. 

\par Recent progress in solar physics instrumentation has revived
the interest in investigating sunspot oscillations. Multiwavelength studies of
the 3 and 5 min waves are of main concern. Whereas at the photosphere-chromosphere
 level these waves are detectable in
intensity and Doppler velocity signals, in the transition zone and the corona they
 are recorded in UV intensity. 


\citet{aschwanden1999} and \citet{naka1999sci} researched transverse spatial oscillations of coronal
loops, whose footpoints were anchored in the photosphere. The oscillations studied in
171\,\AA\ (TRACE) were triggered by a flare and their mean period was found to be 280\,s.
\citet{demoortel2002} suggested
a relation between the 3 min oscillations observed
in coronal loops above sunspots and the 5 min oscillations in neighboring loops.
These authors interpret the 3 min oscillations as slow magnetoacoustic waves
with a propagation velocity of 70--235\,$\mathrm{km\,s^{-1}}$. \citet{bry2003}
also identified them as upwardly traveling acoustic waves.

\par \cite{oshea2002} measured time delays for the 3 min waves detected
at different heights and estimated the propagating speeds for TRACE\,1700 and O\,\textsc{iii}~599.6\,\AA\
to be in the range of 27\hbox{--}86\,$\mathrm{km\,s^{-1}}$. They detected both upward
and downward propagating waves. The result appreciably depends on the slit
position (pixel number). This significantly complicates interpretation of the
time delays detected for the waves at different levels. \citet{marsh2006ApJ} showed
that the 3 min oscillations in the transition zone above sunspots are directly
connected to the waves in coronal loops. They supposed that global \textit{p}-mode
oscillations are involved in the process.

\par There are controversial opinions on plume- and fan-structure connection with  the
3 min umbral oscillations. \citet{Brynildsen2004} showed that the 3 min umbral oscillations
are limited to small coronal regions coinciding with bases of umbral coronal loops. They
argue that these oscillations are not connected to plumes, whereas \citet{jess2012ApJ}
found a direct connection of the 3 min oscillations to coronal fans. They showed that the
sources of the 3 min oscillations in coronal fans are seen to anchor
into photospheric umbral dots with enhanced oscillation power. Analyzing  the 3 min
oscillations, \citet{rez2012} obtained contradictory results for two different sunspots. \citet{Wang2009}
detected 12 and 25 min oscillations of intensity and line-of-sight (LOS) velocity in fan-like coronal
structures above active regions in Hinode Extreme-Ultraviolet Imaging Spectrometer (EIS) data. They identified them as propagating slow
magnetoacoustic waves.

\par The 3 min oscillations detected in the microwave range were found to replicate those recorded
in the chromosphere of sunspots with a 50-second delay \citep{abrmax2011}.
This implies that the waves propagate from the chromosphere of an umbra to the upper levels.

\par Attempts to find a direct connection between photospheric--chromospheric waves and
coronal waves above sunspots have encountered difficulties. As a rule, instruments
that obtain UV data to detect coronal waves do not observe in the H$\alpha$ and
He\,\textsc{i}~10830\,\AA\ lines that are well suited for such a task.
To perform a joint analysis, one has to use data from ground-based
telescopes. The connection between waves in the corona and the deeper levels
can be searched by direct comparison of the signal variations and by studying
 their spatial and temporal properties both visually and using cross-correlation.

\par This paper presents an attempt to make a joint analysis of the oscillations
in a sunspot detected by ground-based observations and the Solar Dynamics
Observatory (SDO) simultaneously. The data cover several levels of the solar
atmosphere from the temperature minimum to the corona. Spatial power
distribution of different oscillations is of interest, as well as their time
lags. We tried to identify the frequencies for which the power spatial
distribution better reproduces fan structures in the 171\,\AA\ line.

\par The measurements of the time lag between the low-frequency oscillations at the
different heights were performed taking into account the oscillation propagation
trajectory inclination. We hope that this allows us to make more accurate wave propagation speed
calculations.

\section{Method and instruments}

\par
\begin{figure}[t]
  \centering
  \includegraphics[width=5cm]{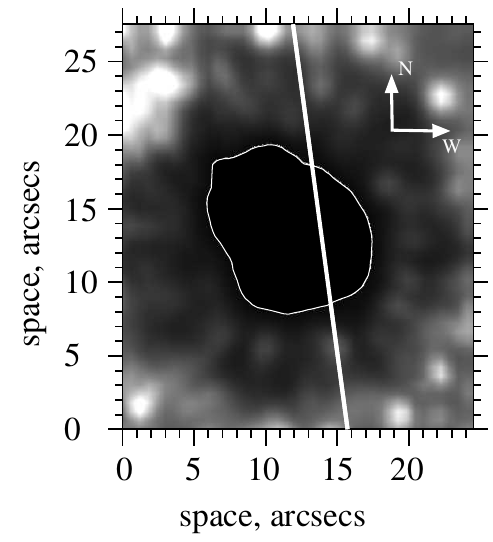}
  \caption{Spot NOAA 11479 in the 1700\,\AA\ continuum. The position of the
slit relative to the spot is shown with the straight white line. The closed
curve denotes the umbra boundary in the white light.}
\label{fig:spot-slit}
\end{figure}

 Two types of data were used in the research. First, we observed the sunspot
NOAA 11479 (Figure~\ref{fig:spot-slit}) close to the disk center on 16\,May\,2012
for eight hours with a ground-based solar telescope at the Sayan Solar Observatory
in two lines simultaneously: Si\,\textsc{i}~10827\,\AA\ and
He\,\textsc{i}~10830\,\AA. They represent the upper photosphere and the chromosphere,
respectively. The coordinates of the spot were N15E15 at the beginning of the
observation, which corresponds to the small angle between the line of sight and
the normal to the Sun's surface. The analysis was performed at different frequency bands, 1--3.5\,mHz and
5--8\,mHz.

\par The telescope resolution is usually about 1$''$ owing to the Earth's
atmosphere. The telescope photoelectric guide tracks the solar image with
1$''$ accuracy for several hours of observations and compensates for image
moving caused by the Sun's rotation.

\par One camera sensor element corresponded to the 0.3$''$ spatial resolution
along the slit and 30\,m\AA\ one along the spectrograph dispersion. The ultimate
cadence was one second after the summation of every ten frames.

\par We used the Doppler-compensator method to calculate the line-of-sight
velocity for the Si\,\textsc{i}--He\,\textsc{i} pair. The wavelength position
shift corresponds to the measured line-of-sight velocity variation
\citep{kob2009ARep}.

\par Also, we used for our analysis a series of Atmospheric Imaging Assembly (AIA) SDO data corresponding to the ground-based
telescope observations. These observations comprise 6173\,\AA, 1700\,\AA, 304\,\AA, and 171\,\AA\
spectral bands, which represent the lower photosphere, the upper
photosphere, the transition region, and the corona, respectively
\citep{rez2012ApJ}.

\par To trace wave propagation from the photosphere to the chromosphere and
the corona, we compared oscillations of the velocity and intensity
signals filtered with Morlet wavelet. The time lags between the signals were
determined by using the cross-correlation method.

\par We based our analysis on the following premises: a) in sunspots the upwardly
propagating waves travel along magnetic field lines; b) registered periodic variations
of line intensities in the lower atmosphere are
caused by temperature variations; c) we use the line formation heights accepted in the current literature for the
spectral lines that we chose.

\begin{figure}[b]
  \centering
  \includegraphics[width=10cm]{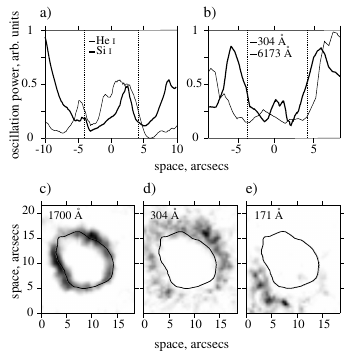}
  \caption{Spatial distribution of the 5 min oscillation power a) along the slit
of the ground-based telescope (Si\,\textsc{i}~10827\,\AA\ and
He\,\textsc{i}~10830\,\AA); b) along the slit in the 304\,\AA\ and HMI Fe\,\textsc{i}~6173\,\AA\
signals; c) in the 1700\,\AA\ continuum; d) in the 304\,\AA\
line; e) in the 171\,\AA\ line. Only the pixels with oscillation power
exceeding 3$\sigma^2$ level are plotted on the distributions on the panels c--e.
The darker a point, the higher is the oscillation power in it.}
  \label{fig:distr_5-min}
\end{figure}

\section{Results}

\par Spatial localization of different frequency oscillations is in close
relation to the magnetic field topology in sunspots. Both the early works
\citep{SigwarthMattig,kob2004,tziotziou2006} and recent publications based
on SDO data analysis \citep{rez2012} note that low-frequency
oscillations concentrate in regions where the magnetic field lines are
significantly inclined (i.e., in the penumbra), whereas the 5\,mHz and higher frequency
oscillations are concentrated within the umbra boundaries.

\par First of all, we computed spatial distribution of the Fourier oscillation
power for the 0.8\,mHz spectral bandpass centered at 3.3\,mHz
(Figure~\ref{fig:distr_5-min}). Only the pixels with oscillation power exceeding
3$\sigma^2$ level are plotted on the power spectrum maps (c-e). The darker a point, the
higher is the oscillation power in it within the specified band pass. The closed curves
denote the umbra border.

\begin{figure}[t]
  \centering
  \includegraphics[width=11cm]{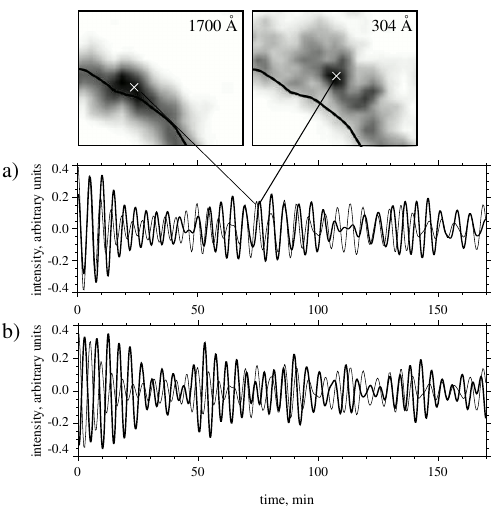}
  \caption{Phase relations between the 5 min oscillations at 1700\,\AA\
(thin) and 304\,\AA\ (thick) levels. The signals were filtered in the 0.8 mHz
band centered at 3.3 mHz. a) The points for the different layers were selected
accounting for the trajectory inclination, so that the filtered oscillation
wave trains in the points showed the maximal correspondence (correlation coefficient is 0.7).
 b) The signals from the points located above each other at the line of sight at the point
  marked in the 1700\,\AA\ panel show less correspondence (correlation coefficient is 0.35).}
  \label{fig:trains_for_1700n304_nonLOS_LOS}
\end{figure}

\par
\begin{figure}[t]
  \centering
  \includegraphics[width=10cm]{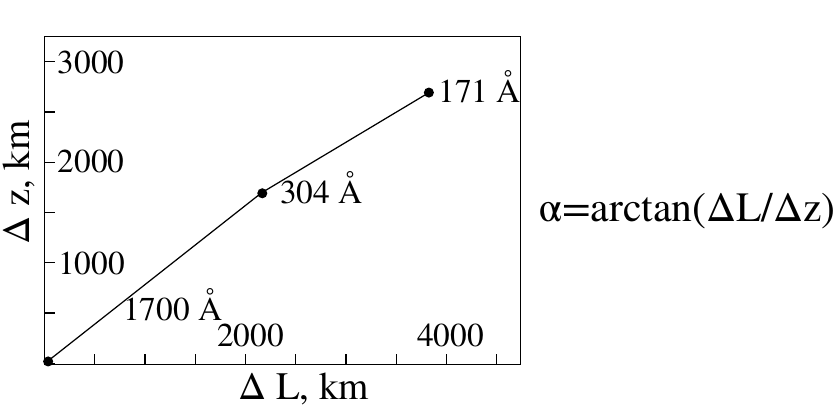}
  \caption{Graphical interpretation of the 5 min wave trajectory. The broken
line represents the average magnetic field inclination to the vertical derived
from the horizontal displacements ($\Delta$L) and the
estimations of the differences between the line formation heights ($\Delta$z).}
  \label{fig:inclination}
\end{figure}

 One can clearly see that the oscillation power is predominantly located in
circle-shape structures; the areas of the circles increase with the line
formation height. Proceeding from the assumption that these oscillations are
manifestations of the magnetohydrodynamic waves propagating along the magnetic field lines, we
attempted to measure the inclination of the lines by
visually matching the oscillation power maps in different bands. To determine
more accurate spatial correspondence, we analyzed sequences of the filtered
signal wave trains in the vicinities of the supposed correspondence points and
found correlation coefficients for them
(Figure~\ref{fig:trains_for_1700n304_nonLOS_LOS}a). Unambiguous correspondence was
not found in every section along the umbra border, but in some points the
cross-correlation coefficients reach 0.75. The mean displacement $\Delta$L of
 the circle structure boundaries is 2150\,km (from
1800 to 3200 for different segments along the umbra boundary) for the
1700\,\AA--304\,\AA\ pair of lines. A minor discrepancy in positioning can make a
contribution to the value spread for this pair of lines. The mean displacement
of the circle boundaries for the 304\,\AA--171\,\AA\ pair of lines is about
1700\,km. The propagation trajectory inclination angle $\alpha$ was approximately
estimated using a formula $\alpha$=arctan($\Delta$L/$\Delta$z), where $\Delta$z
is the formation level difference. To calculate phase velocities, the
propagation trajectory length  S is roughly derived from a formula S=$\Delta$L/sin$\alpha$.
According to our preliminary results, which were calculated taking into account the
differences in line formation heights, the average inclination of the magnetic
field lines is 50-55 and 55-60 degrees for the height ranges corresponding to
the 1700\,\AA--304\,\AA\ and 304\,\AA--171\,\AA\ pairs of lines, respectively
(Figure~\ref{fig:inclination}). Our estimations for the magnetic field line
inclination at the umbra boundary are in compliance with the recent results from
\citet{jess2012ApJ} and \citet{rez2012}.

\par Direct measurements of the signal phase difference at the two height levels
are hindered owing to such an extreme inclination. Wave-train structures
obtained at the line of sight for the two levels do not correspond, and an
unambiguous determination of the phase lag turns out to be impossible
(Figure~\ref{fig:trains_for_1700n304_nonLOS_LOS}b). However, if we compare the signals
in the points chosen with the account for the trajectory inclination, the
similarity of the wave-train structures increases significantly
(Figure~\ref{fig:trains_for_1700n304_nonLOS_LOS}a). Nevertheless, different wave
trains show different phase lags. This can be caused by the uncertainties of
spatial alignment.

\par  The inclination angle is harder to estimate from the
Si\,\textsc{i}--He\,\textsc{i} pair
because we are dealing with a one-dimensional cut of the spot by the
spectrograph entrance slit; thus we can trace only displacements along the slit.
The wave-train correspondence between the two pairs of points that we had is poor.

\par The distributions of Si\,\textsc{i} and He\,\textsc{i} 5 min oscillation
power along the slit within the umbra are peaks with maxima located close to the
umbra center and minima at the umbra borders (Figure~\ref{fig:distr_5-min}a). The
He\,\textsc{i} peak is visibly broader than the Si\,\textsc{i} peak. The full
width at half maximum of the Si\,\textsc{i} peak is 1.6 times smaller than that
of the He\,\textsc{i} peak.

\par The 5 min oscillations in the 1700\,\AA\ and 304\,\AA\ lines are
registered within the umbra, although it is not seen from the greyscale figures
because the oscillation power in the penumbra is much higher than that in the umbra.
The cuts along the slit of the Fe\,\textsc{i}~6173\,\AA\ (HMI)
and He\,\textsc{ii}~304\,\AA\ 5 min oscillation
power maps (Figure~\ref{fig:distr_5-min}) show the presence of peaks in
the umbra, similar to those seen in the ground-based
telescope observations. The displacement between the different layer umbral
peaks is close to that observed for the Si\,\textsc{i}--He\,\textsc{i} pair.

\par
\begin{figure}[t]
  \centering
  \includegraphics[width=13cm]{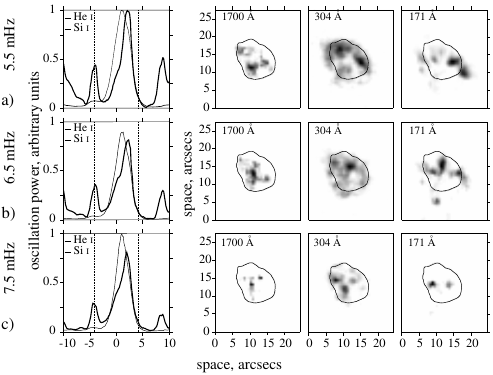}
  \caption{Oscillation power spatial distribution for the bands 0.8\,mHz centered
at a)\,5.5\,mHz; b)\,6.5\,mHz; c)\,7.5\,mHz. The first column presents the power
distribution along the entrance slit for the Si\,\textsc{i} and He\,\textsc{i}
lines.}
  \label{fig:higher_frqncs_dstrbtns}
\end{figure}

 The full width at
the peak half maximum of the oscillation power
distribution in the He\,\textsc{i} line exceeds that of the Si\,\textsc{i} line
in the other frequency bands (5--8\,mHz) as well
(Figure~\ref{fig:higher_frqncs_dstrbtns}). The ratio is 1.6 for the 5.5\,mHz and
decreases to 1 with frequency increasing to 7.5\,mHz. There is a 2$''$ shift
between the Si\,\textsc{i} and He\,\textsc{i} peaks in the oscillation power distributions
(Figure~\ref{fig:higher_frqncs_dstrbtns}, first column). The positional angle
does not cause such a displacement because the spot was observed close to the
disk center. In our opinion, the displacement is caused by the magnetic field inclination.

\par
\begin{figure}[t]
  \centering
  \includegraphics[width=12cm]{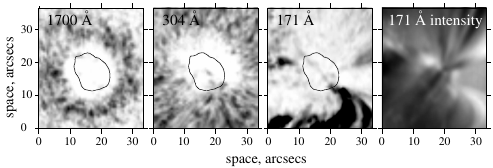}
  \caption{Propagation of the low-frequency oscillations. First three panels
from left to right: spatial distribution of the 15 min oscillation power in the
1700\,\AA, 304\,\AA, and 171\,\AA. The fourth panel shows an image of the
region in the 171\,\AA\ line intensity.}
  \label{fig:distr_15-min}
\end{figure}

 Spatial power distributions for the lower frequency oscillations are
presented in Figure~\ref{fig:distr_15-min}. As one can see, the 15 min
oscillations are concentrated in the outer penumbra in the higher photosphere
(1700\,\AA), forming a circle structure. This structure expands in the transition
zone (304\,\AA). Slender elements stretching radially become evident at this
level. In the end, the 15 min oscillation power distribution in the
171\,\AA\ line partially reproduces the fan picture seen in the corona. This fact
gives evidence that the low-frequency oscillations of the 10--15 min periods
propagate upward from the outer penumbra and reach the corona level. It is of
interest that previously \citet{2009ApJ...698..397V} found transverse 11 min
oscillations of the coronal loops above active regions, which they identified as a fast magnetoacoustic kink mode.

\par Spatial distribution of the high-frequency oscillations (5.5--7.5\,mHz)
differs from the distribution of the low-frequency oscillations
(Figure~\ref{fig:higher_frqncs_dstrbtns}). First, the main power is concentrated
within the umbra at the upper photospheric and chromospheric levels. Second, although
the oscillation region expands significantly at the 304\,\AA\ (transition region)
and the 171\,\AA\ (lower corona) heights, no circle structure appears, but
fragmentariness becomes more apparent. Third, the number of umbral fragments
showing presence of the oscillations decreases with increasing frequency. This
means that the high-frequency oscillations concentrate in special points of the
umbra. According to the finding of \citet{jess2012ApJ}, these points are the chromospheric umbral dots with the increased oscillation power.

\par
\begin{figure}[t]
  \centering
  \includegraphics[width=12cm]{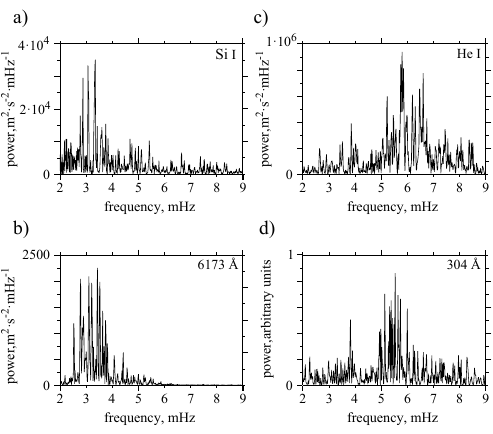}
  \caption{Umbral oscillation power spectra from the photosphere
   to the transition zone for the a) Si\,\textsc{i}~10827\,\AA\ LOS velocity;
  b) Fe\,\textsc{i}~6173\,\AA\ LOS velocity; c) He\,\textsc{i}~10830\,\AA\ LOS
   velocity; d) He\,\textsc{ii}~304\,\AA\ intensity signal.}
\label{fig:spectra}
\end{figure}

 The power spectra (Figure~\ref{fig:spectra}) calculated in the different
lines for the region that is cut by the slit in the sunspot umbra show that the main changes
occur at the heights from the temperature minimum (300-500 km) to the transition region
(2000-2500 km). Five-minute oscillations dominate explicitly in the Fe\,\textsc{i}~6173\,\AA\ (250 km) and
Si\,\textsc{i}~10827\,\AA\ (540 km) LOS velocity power spectra, whereas frequencies from 5.2 to 7 mHz
dominate at the He\textsc{i} 10830\,\AA\ (2100 km) and 304\,\AA\ (2300 km) line formation heights. This
indicates that the 3 min
umbral oscillations evolve and sharply amplify directly in a chromosphere
cavity \citep{1985A&A...143..201Z, Botha2011}. We obtained similar  results earlier from
H$\alpha$ and Fe\,\textsc{i}~6569\,\AA\ observations \citep{kob2011}.

\par Taking into account the horizontal shift of the oscillation between the
layers, we measured time lags for the 5 min oscillations. For the
1700\,\AA--304\,\AA\ pair of lines, the value of the time lag shows a great spread
from -20\,s to 108\,s, where negative values signify the photospheric signal
lagging behind the chromospheric one. A significant spread can appear even in
one spatial point at different time intervals. \citet{rez2012ApJ} note
the similar feature of time lag measurements between the oscillations above
sunspots. The average lag between the Si\,\textsc{i} and He\,\textsc{i} signals
is 47\,s; it shows a great spread, too. The 304\,\AA--171\,\AA\ pair shows the time
lags from -12 to 48\,s, with the average value being 24\,s. Poor correspondence
between the signal wave trains in the layers does not mean that there are no
waves propagating through the layers upward. This may signify the presence of
both standing and traveling waves in the volume researched. The
presence of standing waves in an object does not exclude the
possibility of this object being a source of waves propagating in the
surrounding medium. If the resonator boundary does not reflect the wave
thoroughly, this makes it an oscillation source. The majority of
oscillation sources contain resonators. The ratio between
these two types of waves influences the results obtained. (see, e.g.,
\citet{kob2011}). Also, this may signify that the magnetic field
loops presenting in the lower layer either do not reach the upper level
returning to the photosphere or divert from the line of sight significantly.

\par
\begin{figure}[t]
  \centering
  \includegraphics[width=12cm]{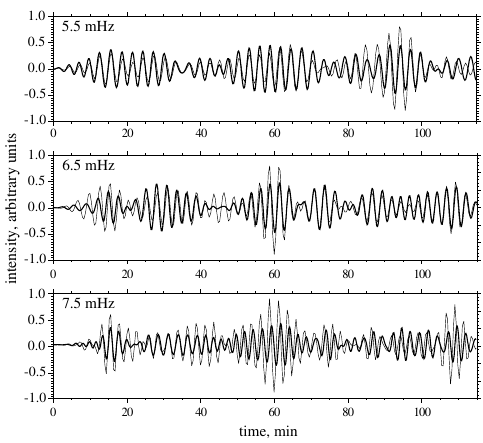}
  \caption{Signals of the 1700\,\AA\ (thin) and 304\,\AA\ (thick) lines filtered
in 0.8\,mHz band centered at 5.5\,mHz, 6.5\,mHz, and 7.5\,mHz. The 304\,\AA\ signal
 was shifted backwards by \textit{$\delta$t}, so that the correlation coefficients between the signals
  were maximal ($\delta$t is 36\,s, 60\,s, and 52\,s for the upper, middle, and bottom panel, respectively).
   The 1700\,\AA\ signal was amplified by a factor of 3.3, 5, and 6.7 for the first, second, and third band, respectively.}
\label{fig:1700thin_and_304thick_08_55_mHz}
\end{figure}

\begin{figure}[t]
  \centering
  \includegraphics[width=12cm]{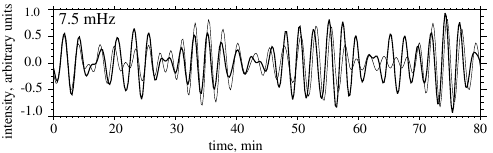}
  \caption{Unshifted signals of the 304\,\AA\ (thick) and 171\,\AA\ (thin) lines filtered
in 0.8\,mHz band centered at 7.5\,mHz. The 171\,\AA\ signal was amplified
 by a factor of 2.3.}
\label{fig:304thin_and_171thick_0.8_7.5_mHz}
\end{figure}

 The time lag varies from 36 to 84\,s and reaches 0 at some spatial points
for the 1700\,\AA--304\,\AA\ pair in the 0.8\,mHz band passes centered at 5.5, 6.5,
7.5\,mHz (see Figure~\ref{fig:1700thin_and_304thick_08_55_mHz}). We chose the
points located above each other at the line of sight for our analysis. Because the high-frequency oscillations propagate along the vertical
and close-to-vertical magnetic field lines in
the central umbra, the visibility of the traveling waves is hardly influenced
by a projection effect. The wave-train
structure resemblance indicates that the points located at the oscillation
propagation trajectory are used. Besides, wave-train visual analysis helps avoid
the 2$\pi$ uncertainty, which
occurs when one measures phase lags. In the
304\,\AA--171\,\AA\ pair, the lag varies from 12 to 24 s (see, e.g., Figure~\ref{fig:304thin_and_171thick_0.8_7.5_mHz}). One can see that the wave
trains in
Figures~\ref{fig:1700thin_and_304thick_08_55_mHz} and~\ref{fig:304thin_and_171thick_0.8_7.5_mHz} correspond to each other much better
than those of the lower frequencies. This is because the
higher frequencies `prefer' to propagate along the vertical magnetic flux.

\par Taking the height difference of 1700\,km for the 1700\,\AA--304\,\AA\
pair and 1000\,km for the 304\,\AA--171\,\AA\ pair
\citep{rez2012ApJ} and bearing in mind the diversion from the vertical propagation,
we found the average phase velocities to be
24\,$\mathrm{km\,s^{-1}}$ and 42\,$\mathrm{km\,s^{-1}}$ for the first and the
second pair, respectively, for the frequency band centered at 3.3\,mHz. For the
frequencies from 5 to 8\,mHz, the phase velocities are 28\,$\mathrm{km\,s^{-1}}$
and 55\,$\mathrm{km\,s^{-1}}$, respectively. The uncertainties of these
calculations are relatively great (from 20\% to 60\%) owing to the
inconsistencies in the determined time lags.

\par It is of interest to note that for all the frequencies analyzed, the
 304\,\AA\ line intensity oscillation amplitude exceeds appreciably those of
  the other lines, both at the higher and lower levels (see also \citet{rez2012ApJ}).

\par
\begin{figure}[t]
  \centering
  \includegraphics[width=12cm]{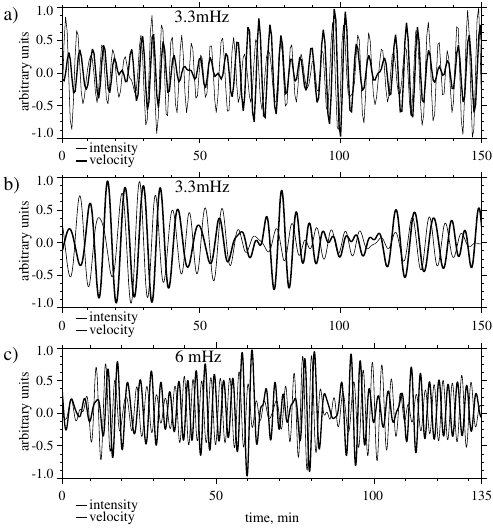}
  \caption{Phase relationships between the intensity (thin) and velocity (thick)
  signals for the a) Fe\,\textsc{i}~6173\,\AA\ line; b) Si\,\textsc{i}~10827\,\AA\
  line; c) He\,\textsc{i}~10830\,\AA\ line.}
\label{fig:IV}
\end{figure}

 We found that the phase difference between the intensity and velocity signals
(I-V) for the Fe\,\textsc{i}~6173\,\AA\
and Si\,\textsc{i}~10827\,\AA\ lines is close to $90^{\circ}$ on average (Figure~\ref{fig:IV}), which is
characteristic for acoustic waves under the adiabatic approximation. This difference
is ambiguous and close to $180^{\circ}$ for the upper chromosphere
He\,\textsc{i}~10830\,\AA\ line at individual
time intervals. The situation with the He\,\textsc{i}~10830\,\AA\
line is complicated by the fact that its
profile is largely influenced by the UV radiation coming from the upper layers
of the solar atmosphere. We assume that the observed waves are slow magnetoacoustic
waves propagating upward
along the magnetic field lines.

\par We believe that intensity-temperature phase relationship research in
the corona is also important. To this end, we intend to implement the
promising method described in \citet{AschwandenBoe2011} and \citet{Aschwan2013}
in our future work.

\section{Conclusions}

\par The spatial distribution of 5 min oscillation power looks like a
circle-shape structure matching the umbra border at the photospheric level.
The circle expands, and its boundaries move away from the inner penumbra border
at the higher levels (304\,\AA\ and 171\,\AA). This indicates that 5 min
oscillations propagate along appreciably inclined magnetic tubes in the spot.
The inclination of the tubes reaches 50--60 degrees at the levels of the
transition region and the lower corona.

\par While the 10--15 min oscillation power distribution shows a circle structure at
the lower layers as well, it clearly resembles fan structures at the coronal level (it is of interest to notice that the coronal-level 5 min distribution
does not show such a resemblance, at least up to the 171\,\AA\ formation
level). As is known, 8-15 min oscillations dominate in the photospheric and chromospheric outer penumbra (see \citet{SigwarthMattig,Kobanov2000SoPh}).
  This indicates that these oscillations come to the corona from the lower layers
of the solar atmosphere.

\par The main oscillation power in the 5--8\,mHz range is concentrated within
the umbra boundaries both at the temperature-minimum and chromospheric levels. The
oscillation region covers the greater area at the transition region and at lower
corona levels than that at the photospheric level. The higher frequency
oscillations are concentrated in small regions inside the umbra, probably
coinciding with the loop bases anchored in umbral dots.

\par We found no evidence of any connection between the 3 min oscillations and fan structures, at least for the 171\,\AA\ formation
level. In this respect our results agree with
those of \citet{Brynildsen2004}. It is reasonable to suppose that  the connection
between the 3 min oscillations and fan structures is more evident in the higher
coronal levels.

\par We deduced that the average phase velocities are 26$\pm$15
$\mathrm{km\,s^{-1}}$ and 50$\pm$10 $\mathrm{km\,s^{-1}}$ in the frequency range
3--8 mHz for the 1700\,\AA--304\,\AA\ and 304\,\AA--171\,\AA\ height levels,
respectively.

\par To determine the phase lags of the waves between the height levels more
correctly, it is necessary to take into account the inclination of the magnetic
field lines, i.e., to analyze the points not taken at the line of sight but
rather chosen considering the displacement in the horizontal plane. As a result,
the direct phase-lag detection procedure becomes considerably more complicated.
This especially applies to the low-frequency oscillations.
\vskip1cm

\small{\textbf{Acknowledgements}. This study was supported in part by the RFBR
research project No.: 12\hbox{-}02\hbox{-}33110 mol\_a\_ved \, and \,
the Grant of the President of the Russian Federation
No.: MK\hbox{-}497.2012.2, Russian Federation
Ministry of Education and Science state contract No.: 14.518.11.7047 and
agreement No.: 8407. We acknowledge NASA/SDO science team for providing the data.
We acknowledge Y.M.~Kaplunenko for his help in preparing the English version of the paper.
}

\bibliographystyle{spr-mp-sola}
\bibliography{kobanov-2012}

\begin{thebibliography}{35}
\ifx \bisbn   \undefined \def \bisbn  #1{ISBN #1}\fi
\ifx \binits  \undefined \def \binits#1{#1}\fi
\ifx \bauthor  \undefined \def \bauthor#1{#1}\fi
\ifx \batitle  \undefined \def \batitle#1{#1}\fi
\ifx \bjtitle  \undefined \def \bjtitle#1{\textit{#1}}\fi
\ifx \bvolume  \undefined \def \bvolume#1{\textbf{#1}}\fi
\ifx \byear  \undefined \def \byear#1{#1}\fi
\ifx \bissue  \undefined \def \bissue#1{#1}\fi
\ifx \bfpage  \undefined \def \bfpage#1{#1}\fi
\ifx \blpage  \undefined \def \blpage #1{#1}\fi
\ifx \burl  \undefined \def \burl#1{\textsf{#1}}\fi
\ifx \href  \undefined \def \href#1#2{\textsf{#2}}\fi
\ifx \doiurl  \undefined \def
  \doiurl#1{\href{http://dx.doi.org/#1}{\textsf{#1}}}\fi
\ifx \betal  \undefined \def \betal{\textit{et al.}}\fi
\ifx \binstitute  \undefined \def \binstitute#1{#1}\fi
\ifx \bctitle  \undefined \def \bctitle#1{#1}\fi
\ifx \beditor  \undefined \def \beditor#1{#1}\fi
\ifx \bpublisher  \undefined \def \bpublisher#1{#1}\fi
\ifx \bbtitle  \undefined \def \bbtitle#1{\textit{#1}}\fi
\ifx \bedition  \undefined \def \bedition#1{#1}\fi
\ifx \bseriesno  \undefined \def \bseriesno#1{\textbf{#1}}\fi
\ifx \blocation  \undefined \def \blocation#1{#1}\fi
\ifx \bsertitle  \undefined \def \bsertitle#1{\textit{#1}}\fi
\ifx \bsnm \undefined \def \bsnm#1{#1}\fi
\ifx \bsuffix \undefined \def \bsuffix#1{#1}\fi
\ifx \bparticle \undefined \def \bparticle#1{#1}\fi
\ifx \barticle \undefined \def \barticle#1{}\fi
\ifx \botherref \undefined \def \botherref#1{}\fi
\ifx \url \undefined \def \url#1{\textsf{#1}}\fi
\ifx \bchapter \undefined \def \bchapter#1{}\fi
\ifx \bbook \undefined \def \bbook#1{}\fi
\ifx \bcomment \undefined \def \bcomment#1{#1}\fi
\ifx \oauthor \undefined \def \oauthor#1{#1}\fi
\ifx \citeauthoryear \undefined \def \citeauthoryear#1{#1}\fi
\def \endbibitem {}
\ifx \bconflocation  \undefined \def \bconflocation#1{#1} \fi

\bibitem[\protect\citeauthoryear{{Abramov-Maximov}
  \textit{et~al.}}{2011}]{abrmax2011}
\begin{barticle}
\bauthor{\bsnm{{Abramov-Maximov}}, \binits{V.E.}},
\bauthor{\bsnm{{Gelfreikh}}, \binits{G.B.}},
\bauthor{\bsnm{{Kobanov}}, \binits{N.I.}},
\bauthor{\bsnm{{Shibasaki}}, \binits{K.}},
\bauthor{\bsnm{{Chupin}}, \binits{S.A.}}:
\byear{2011},
\batitle{{Multilevel Analysis of Oscillation Motions in Active Regions of the
  Sun}}.
\bjtitle{\solphys}
\bvolume{270},
\bfpage{175}\,--\,\blpage{189}.
doi:\doiurl{10.1007/s11207-011-9716-7}.
\end{barticle}
\endbibitem

\bibitem[\protect\citeauthoryear{{Aschwanden} and
  {Boerner}}{2011}]{AschwandenBoe2011}
\begin{barticle}
\bauthor{\bsnm{{Aschwanden}}, \binits{M.J.}},
\bauthor{\bsnm{{Boerner}}, \binits{P.}}:
\byear{2011},
\batitle{{Solar Corona Loop Studies with the Atmospheric Imaging Assembly. I.
  Cross-sectional Temperature Structure}}.
\bjtitle{\apj}
\bvolume{732},
\bfpage{81}.
doi:\doiurl{10.1088/0004-637X/732/2/81}.
\end{barticle}
\endbibitem

\bibitem[\protect\citeauthoryear{{Aschwanden}
  \textit{et~al.}}{1999}]{aschwanden1999}
\begin{barticle}
\bauthor{\bsnm{{Aschwanden}}, \binits{M.J.}},
\bauthor{\bsnm{{Fletcher}}, \binits{L.}},
\bauthor{\bsnm{{Schrijver}}, \binits{C.J.}},
\bauthor{\bsnm{{Alexander}}, \binits{D.}}:
\byear{1999},
\batitle{{Coronal Loop Oscillations Observed with the Transition Region and
  Coronal Explorer}}.
\bjtitle{\apj}
\bvolume{520},
\bfpage{880}\,--\,\blpage{894}.
doi:\doiurl{10.1086/307502}.
\end{barticle}
\endbibitem

\bibitem[\protect\citeauthoryear{{Aschwanden}
  \textit{et~al.}}{2013}]{Aschwan2013}
\begin{barticle}
\bauthor{\bsnm{{Aschwanden}}, \binits{M.J.}},
\bauthor{\bsnm{{Boerner}}, \binits{P.}},
\bauthor{\bsnm{{Schrijver}}, \binits{C.J.}},
\bauthor{\bsnm{{Malanushenko}}, \binits{A.}}:
\byear{2013},
\batitle{{Automated Temperature and Emission Measure Analysis of Coronal Loops
  and Active Regions Observed with the Atmospheric Imaging Assembly on the
  Solar Dynamics Observatory (SDO/AIA)}}.
\bjtitle{\solphys}
\bvolume{283},
\bfpage{5}\,--\,\blpage{30}.
doi:\doiurl{10.1007/s11207-011-9876-5}.
\end{barticle}
\endbibitem

\bibitem[\protect\citeauthoryear{{Balthasar} and
  {Wiehr}}{1990}]{1990A&A...237..243B}
\begin{barticle}
\bauthor{\bsnm{{Balthasar}}, \binits{H.}},
\bauthor{\bsnm{{Wiehr}}, \binits{E.}}:
\byear{1990},
\batitle{{Oscillations of Evershed velocities and asymmetries}}.
\bjtitle{\aap}
\bvolume{237},
\bfpage{243}\,--\,\blpage{246}.
\end{barticle}
\endbibitem

\bibitem[\protect\citeauthoryear{{Beckers} and {Tallant}}{1969}]{beckers1969uf}
\begin{barticle}
\bauthor{\bsnm{{Beckers}}, \binits{J.M.}},
\bauthor{\bsnm{{Tallant}}, \binits{P.E.}}:
\byear{1969},
\batitle{{Chromospheric Inhomogeneities in Sunspot Umbrae}}.
\bjtitle{\solphys}
\bvolume{7},
\bfpage{351}.
\end{barticle}
\endbibitem

\bibitem[\protect\citeauthoryear{{Bloomfield}, {Lagg}, and
  {Solanki}}{2007}]{bloomfield2007}
\begin{barticle}
\bauthor{\bsnm{{Bloomfield}}, \binits{D.S.}},
\bauthor{\bsnm{{Lagg}}, \binits{A.}},
\bauthor{\bsnm{{Solanki}}, \binits{S.K.}}:
\byear{2007},
\batitle{{The Nature of Running Penumbral Waves Revealed}}.
\bjtitle{\apj}
\bvolume{671},
\bfpage{1005}\,--\,\blpage{1012}.
doi:\doiurl{10.1086/523266}.
\end{barticle}
\endbibitem

\bibitem[\protect\citeauthoryear{{Bogdan} and {Judge}}{2006}]{bogdan2006a}
\begin{barticle}
\bauthor{\bsnm{{Bogdan}}, \binits{T.J.}},
\bauthor{\bsnm{{Judge}}, \binits{P.G.}}:
\byear{2006},
\batitle{{Observational aspects of sunspot oscillations}}.
\bjtitle{Royal Society of London Philosophical Transactions Series A}
\bvolume{364},
\bfpage{313}\,--\,\blpage{331}.
\end{barticle}
\endbibitem

\bibitem[\protect\citeauthoryear{{Botha} \textit{et~al.}}{2011}]{Botha2011}
\begin{barticle}
\bauthor{\bsnm{{Botha}}, \binits{G.J.J.}},
\bauthor{\bsnm{{Arber}}, \binits{T.D.}},
\bauthor{\bsnm{{Nakariakov}}, \binits{V.M.}},
\bauthor{\bsnm{{Zhugzhda}}, \binits{Y.D.}}:
\byear{2011},
\batitle{{Chromospheric Resonances above Sunspot Umbrae}}.
\bjtitle{\apj}
\bvolume{728},
\bfpage{84}.
doi:\doiurl{10.1088/0004-637X/728/2/84}.
\end{barticle}
\endbibitem

\bibitem[\protect\citeauthoryear{{Brynildsen} \textit{et~al.}}{2003}]{bry2003}
\begin{barticle}
\bauthor{\bsnm{{Brynildsen}}, \binits{N.}},
\bauthor{\bsnm{{Maltby}}, \binits{P.}},
\bauthor{\bsnm{{Brekke}}, \binits{P.}},
\bauthor{\bsnm{{Redvik}}, \binits{T.}},
\bauthor{\bsnm{{Kjeldseth-Moe}}, \binits{O.}}:
\byear{2003},
\batitle{{Search for a chromospheric resonator above sunspots}}.
\bjtitle{Advances in Space Research}
\bvolume{32},
\bfpage{1097}\,--\,\blpage{1102}.
doi:\doiurl{10.1016/S0273-1177(03)00312-0}.
\end{barticle}
\endbibitem

\bibitem[\protect\citeauthoryear{{Brynildsen}
  \textit{et~al.}}{2004}]{Brynildsen2004}
\begin{barticle}
\bauthor{\bsnm{{Brynildsen}}, \binits{N.}},
\bauthor{\bsnm{{Maltby}}, \binits{P.}},
\bauthor{\bsnm{{Foley}}, \binits{C.R.}},
\bauthor{\bsnm{{Fredvik}}, \binits{T.}},
\bauthor{\bsnm{{Kjeldseth-Moe}}, \binits{O.}}:
\byear{2004},
\batitle{{Oscillations in the Umbral Atmosphere}}.
\bjtitle{\solphys}
\bvolume{221},
\bfpage{237}\,--\,\blpage{260}.
doi:\doiurl{10.1023/B:SOLA.0000035065.10112.fc}.
\end{barticle}
\endbibitem

\bibitem[\protect\citeauthoryear{{Centeno}, {Collados}, and {Trujillo
  Bueno}}{2009}]{centeno2009}
\begin{barticle}
\bauthor{\bsnm{{Centeno}}, \binits{R.}},
\bauthor{\bsnm{{Collados}}, \binits{M.}},
\bauthor{\bsnm{{Trujillo Bueno}}, \binits{J.}}:
\byear{2009},
\batitle{{Wave Propagation and Shock Formation in Different Magnetic
  Structures}}.
\bjtitle{\apj}
\bvolume{692},
\bfpage{1211}\,--\,\blpage{1220}.
doi:\doiurl{10.1088/0004-637X/692/2/1211}.
\end{barticle}
\endbibitem

\bibitem[\protect\citeauthoryear{{De Moortel}
  \textit{et~al.}}{2002}]{demoortel2002}
\begin{barticle}
\bauthor{\bsnm{{De Moortel}}, \binits{I.}},
\bauthor{\bsnm{{Ireland}}, \binits{J.}},
\bauthor{\bsnm{{Hood}}, \binits{A.W.}},
\bauthor{\bsnm{{Walsh}}, \binits{R.W.}}:
\byear{2002},
\batitle{{The detection of 3 \& 5 min period oscillations in coronal loops}}.
\bjtitle{\aap}
\bvolume{387},
\bfpage{L13}\,--\,\blpage{L16}.
doi:\doiurl{10.1051/0004-6361:20020436}.
\end{barticle}
\endbibitem

\bibitem[\protect\citeauthoryear{{Giovanelli}}{1972}]{giovanelli1972}
\begin{barticle}
\bauthor{\bsnm{{Giovanelli}}, \binits{R.G.}}:
\byear{1972},
\batitle{{Oscillations and Waves in a Sunspot}}.
\bjtitle{\solphys}
\bvolume{27},
\bfpage{71}.
\end{barticle}
\endbibitem

\bibitem[\protect\citeauthoryear{{Jess} \textit{et~al.}}{2012}]{jess2012ApJ}
\begin{barticle}
\bauthor{\bsnm{{Jess}}, \binits{D.B.}},
\bauthor{\bsnm{{De Moortel}}, \binits{I.}},
\bauthor{\bsnm{{Mathioudakis}}, \binits{M.}},
\bauthor{\bsnm{{Christian}}, \binits{D.J.}},
\bauthor{\bsnm{{Reardon}}, \binits{K.P.}},
\bauthor{\bsnm{{Keys}}, \binits{P.H.}},
\bauthor{\bsnm{{Keenan}}, \binits{F.P.}}:
\byear{2012},
\batitle{{The Source of 3 Minute Magnetoacoustic Oscillations in Coronal
  Fans}}.
\bjtitle{\apj}
\bvolume{757},
\bfpage{160}.
doi:\doiurl{10.1088/0004-637X/757/2/160}.
\end{barticle}
\endbibitem

\bibitem[\protect\citeauthoryear{{Kobanov}}{2000}]{Kobanov2000SoPh}
\begin{barticle}
\bauthor{\bsnm{{Kobanov}}, \binits{N.I.}}:
\byear{2000},
\batitle{{The properties of velocity oscillations in vicinities of sunspot
  penumbra}}.
\bjtitle{\solphys}
\bvolume{196},
\bfpage{129}\,--\,\blpage{135}.
\end{barticle}
\endbibitem

\bibitem[\protect\citeauthoryear{{Kobanov} and {Makarchik}}{2004}]{kob2004}
\begin{barticle}
\bauthor{\bsnm{{Kobanov}}, \binits{N.I.}},
\bauthor{\bsnm{{Makarchik}}, \binits{D.V.}}:
\byear{2004},
\batitle{{Propagating waves in the sunspot umbra chromosphere}}.
\bjtitle{\aap}
\bvolume{424},
\bfpage{671}\,--\,\blpage{675}.
doi:\doiurl{10.1051/0004-6361:20035960}.
\end{barticle}
\endbibitem

\bibitem[\protect\citeauthoryear{{Kobanov}, {Kolobov}, and
  {Makarchik}}{2006}]{kob2006}
\begin{barticle}
\bauthor{\bsnm{{Kobanov}}, \binits{N.I.}},
\bauthor{\bsnm{{Kolobov}}, \binits{D.Y.}},
\bauthor{\bsnm{{Makarchik}}, \binits{D.V.}}:
\byear{2006},
\batitle{{Umbral Three-Minute Oscillations and Running Penumbral Waves}}.
\bjtitle{\solphys}
\bvolume{238},
\bfpage{231}\,--\,\blpage{244}.
doi:\doiurl{10.1007/s11207-006-0160-z}.
\end{barticle}
\endbibitem

\bibitem[\protect\citeauthoryear{{Kobanov} \textit{et~al.}}{2009}]{kob2009ARep}
\begin{barticle}
\bauthor{\bsnm{{Kobanov}}, \binits{N.I.}},
\bauthor{\bsnm{{Kolobov}}, \binits{D.Y.}},
\bauthor{\bsnm{{Sklyar}}, \binits{A.A.}},
\bauthor{\bsnm{{Chupin}}, \binits{S.A.}},
\bauthor{\bsnm{{Pulyaev}}, \binits{V.A.}}:
\byear{2009},
\batitle{{Characteristics of oscillatory-wave processes in solar structures
  with various magnetic field topology}}.
\bjtitle{Astronomy Reports}
\bvolume{53},
\bfpage{957}\,--\,\blpage{967}.
doi:\doiurl{10.1134/S1063772909100072}.
\end{barticle}
\endbibitem

\bibitem[\protect\citeauthoryear{{Kobanov} \textit{et~al.}}{2011}]{kob2011}
\begin{barticle}
\bauthor{\bsnm{{Kobanov}}, \binits{N.I.}},
\bauthor{\bsnm{{Kolobov}}, \binits{D.Y.}},
\bauthor{\bsnm{{Chupin}}, \binits{S.A.}},
\bauthor{\bsnm{{Nakariakov}}, \binits{V.M.}}:
\byear{2011},
\batitle{{Height distribution of the power of 3-min oscillations over
  sunspots}}.
\bjtitle{\aap}
\bvolume{525},
\bfpage{A41}.
doi:\doiurl{10.1051/0004-6361/200913533}.
\end{barticle}
\endbibitem

\bibitem[\protect\citeauthoryear{{Lites}}{1992}]{lites1992a}
\begin{bchapter}
\bauthor{\bsnm{{Lites}}, \binits{B.W.}}:
\byear{1992},
\bctitle{{Sunspot oscillations - Observations and implications}}.
In: \beditor{\bsnm{{Thomas}}, \binits{J.H.}},
\beditor{\bsnm{{Weiss}}, \binits{N.O.}} (eds.)
\bbtitle{NATO ASIC Proc. 375: Sunspots. Theory and Observations},
\bfpage{261}\,--\,\blpage{302}.
\end{bchapter}
\endbibitem

\bibitem[\protect\citeauthoryear{{Marsh} and {Walsh}}{2006}]{marsh2006ApJ}
\begin{barticle}
\bauthor{\bsnm{{Marsh}}, \binits{M.S.}},
\bauthor{\bsnm{{Walsh}}, \binits{R.W.}}:
\byear{2006},
\batitle{{p-Mode Propagation through the Transition Region into the Solar
  Corona. I. Observations}}.
\bjtitle{\apj}
\bvolume{643},
\bfpage{540}\,--\,\blpage{548}.
doi:\doiurl{10.1086/501450}.
\end{barticle}
\endbibitem

\bibitem[\protect\citeauthoryear{{Nakariakov}
  \textit{et~al.}}{1999}]{naka1999sci}
\begin{barticle}
\bauthor{\bsnm{{Nakariakov}}, \binits{V.M.}},
\bauthor{\bsnm{{Ofman}}, \binits{L.}},
\bauthor{\bsnm{{Deluca}}, \binits{E.E.}},
\bauthor{\bsnm{{Roberts}}, \binits{B.}},
\bauthor{\bsnm{{Davila}}, \binits{J.M.}}:
\byear{1999},
\batitle{{TRACE observation of damped coronal loop oscillations: Implications
  for coronal heating}}.
\bjtitle{Science}
\bvolume{285},
\bfpage{862}\,--\,\blpage{864}.
doi:\doiurl{10.1126/science.285.5429.862}.
\end{barticle}
\endbibitem

\bibitem[\protect\citeauthoryear{{O'Shea}, {Muglach}, and
  {Fleck}}{2002}]{oshea2002}
\begin{barticle}
\bauthor{\bsnm{{O'Shea}}, \binits{E.}},
\bauthor{\bsnm{{Muglach}}, \binits{K.}},
\bauthor{\bsnm{{Fleck}}, \binits{B.}}:
\byear{2002},
\batitle{{Oscillations above sunspots: Evidence for propagating waves?}}
\bjtitle{\aap}
\bvolume{387},
\bfpage{642}\,--\,\blpage{664}.
doi:\doiurl{10.1051/0004-6361:20020375}.
\end{barticle}
\endbibitem

\bibitem[\protect\citeauthoryear{{Reznikova} and {Shibasaki}}{2012}]{rez2012}
\begin{barticle}
\bauthor{\bsnm{{Reznikova}}, \binits{V.E.}},
\bauthor{\bsnm{{Shibasaki}}, \binits{K.}}:
\byear{2012},
\batitle{{Spatial Structure of Sunspot Oscillations Observed with SDO/AIA}}.
\bjtitle{\apj}
\bvolume{756},
\bfpage{35}.
doi:\doiurl{10.1088/0004-637X/756/1/35}.
\end{barticle}
\endbibitem

\bibitem[\protect\citeauthoryear{{Reznikova}
  \textit{et~al.}}{2012}]{rez2012ApJ}
\begin{barticle}
\bauthor{\bsnm{{Reznikova}}, \binits{V.E.}},
\bauthor{\bsnm{{Shibasaki}}, \binits{K.}},
\bauthor{\bsnm{{Sych}}, \binits{R.A.}},
\bauthor{\bsnm{{Nakariakov}}, \binits{V.M.}}:
\byear{2012},
\batitle{{Three-minute Oscillations above Sunspot Umbra Observed with the Solar
  Dynamics Observatory/Atmospheric Imaging Assembly and Nobeyama
  Radioheliograph}}.
\bjtitle{\apj}
\bvolume{746},
\bfpage{119}.
doi:\doiurl{10.1088/0004-637X/746/2/119}.
\end{barticle}
\endbibitem

\bibitem[\protect\citeauthoryear{{Rouppe van der Voort}
  \textit{et~al.}}{2003}]{roupe2003}
\begin{barticle}
\bauthor{\bsnm{{Rouppe van der Voort}}, \binits{L.H.M.}},
\bauthor{\bsnm{{Rutten}}, \binits{R.J.}},
\bauthor{\bsnm{{S{\"u}tterlin}}, \binits{P.}},
\bauthor{\bsnm{{Sloover}}, \binits{P.J.}},
\bauthor{\bsnm{{Krijger}}, \binits{J.M.}}:
\byear{2003},
\batitle{{La Palma observations of umbral flashes}}.
\bjtitle{\aap}
\bvolume{403},
\bfpage{277}\,--\,\blpage{285}.
doi:\doiurl{10.1051/0004-6361:20030237}.
\end{barticle}
\endbibitem

\bibitem[\protect\citeauthoryear{{Sigwarth} and
  {Mattig}}{1997}]{SigwarthMattig}
\begin{barticle}
\bauthor{\bsnm{{Sigwarth}}, \binits{M.}},
\bauthor{\bsnm{{Mattig}}, \binits{W.}}:
\byear{1997},
\batitle{{Velocity and intensity oscillations in sunspot penumbrae.}}
\bjtitle{\aap}
\bvolume{324},
\bfpage{743}\,--\,\blpage{749}.
\end{barticle}
\endbibitem

\bibitem[\protect\citeauthoryear{{Thomas} and
  {Weiss}}{2008}]{2008sust.book.....T}
\begin{bbook}
\bauthor{\bsnm{{Thomas}}, \binits{J.H.}},
\bauthor{\bsnm{{Weiss}}, \binits{N.O.}}:
\byear{2008},
\bbtitle{Sunspots and starspots},
\bpublisher{Cambridge University Press},
\blocation{Cambridge, UK.}
\bisbn{978-0-521-86003-1}.
\end{bbook}
\endbibitem

\bibitem[\protect\citeauthoryear{{Tsiropoula}, {Alissandrakis}, and
  {Mein}}{2000}]{tsiropoula2000}
\begin{barticle}
\bauthor{\bsnm{{Tsiropoula}}, \binits{G.}},
\bauthor{\bsnm{{Alissandrakis}}, \binits{C.E.}},
\bauthor{\bsnm{{Mein}}, \binits{P.}}:
\byear{2000},
\batitle{{Association of chromospheric sunspot umbral oscillations and running
  penumbral waves. I. Morphological study}}.
\bjtitle{\aap}
\bvolume{355},
\bfpage{375}\,--\,\blpage{380}.
\end{barticle}
\endbibitem

\bibitem[\protect\citeauthoryear{{Tziotziou}
  \textit{et~al.}}{2006}]{tziotziou2006}
\begin{barticle}
\bauthor{\bsnm{{Tziotziou}}, \binits{K.}},
\bauthor{\bsnm{{Tsiropoula}}, \binits{G.}},
\bauthor{\bsnm{{Mein}}, \binits{N.}},
\bauthor{\bsnm{{Mein}}, \binits{P.}}:
\byear{2006},
\batitle{{Observational characteristics and association of umbral oscillations
  and running penumbral waves}}.
\bjtitle{\aap}
\bvolume{456},
\bfpage{689}\,--\,\blpage{695}.
doi:\doiurl{10.1051/0004-6361:20064997}.
\end{barticle}
\endbibitem

\bibitem[\protect\citeauthoryear{{Verwichte}
  \textit{et~al.}}{2009}]{2009ApJ...698..397V}
\begin{barticle}
\bauthor{\bsnm{{Verwichte}}, \binits{E.}},
\bauthor{\bsnm{{Aschwanden}}, \binits{M.J.}},
\bauthor{\bsnm{{Van Doorsselaere}}, \binits{T.}},
\bauthor{\bsnm{{Foullon}}, \binits{C.}},
\bauthor{\bsnm{{Nakariakov}}, \binits{V.M.}}:
\byear{2009},
\batitle{{Seismology of a Large Solar Coronal Loop from EUVI/STEREO
  Observations of its Transverse Oscillation}}.
\bjtitle{\apj}
\bvolume{698},
\bfpage{397}\,--\,\blpage{404}.
doi:\doiurl{10.1088/0004-637X/698/1/397}.
\end{barticle}
\endbibitem

\bibitem[\protect\citeauthoryear{{Wang} \textit{et~al.}}{2009}]{Wang2009}
\begin{barticle}
\bauthor{\bsnm{{Wang}}, \binits{T.J.}},
\bauthor{\bsnm{{Ofman}}, \binits{L.}},
\bauthor{\bsnm{{Davila}}, \binits{J.M.}},
\bauthor{\bsnm{{Mariska}}, \binits{J.T.}}:
\byear{2009},
\batitle{{Hinode/EIS observations of propagating low-frequency slow
  magnetoacoustic waves in fan-like coronal loops}}.
\bjtitle{\aap}
\bvolume{503},
\bfpage{L25}\,--\,\blpage{L28}.
doi:\doiurl{10.1051/0004-6361/200912534}.
\end{barticle}
\endbibitem

\bibitem[\protect\citeauthoryear{{Zhugzhda}, {Locans}, and
  {Staude}}{1985}]{1985A&A...143..201Z}
\begin{barticle}
\bauthor{\bsnm{{Zhugzhda}}, \binits{I.D.}},
\bauthor{\bsnm{{Locans}}, \binits{V.}},
\bauthor{\bsnm{{Staude}}, \binits{J.}}:
\byear{1985},
\batitle{{Oscillations in the chromosphere and transition region above sunspot
  umbrae - A photospheric or a chromospheric resonator?}}
\bjtitle{\aap}
\bvolume{143},
\bfpage{201}\,--\,\blpage{205}.
\end{barticle}
\endbibitem

\bibitem[\protect\citeauthoryear{{Zirin} and {Stein}}{1972}]{zirin1972}
\begin{barticle}
\bauthor{\bsnm{{Zirin}}, \binits{H.}},
\bauthor{\bsnm{{Stein}}, \binits{A.}}:
\byear{1972},
\batitle{{Observations of Running Penumbral Waves}}.
\bjtitle{\apjl}
\bvolume{178},
\bfpage{L85}.
\end{barticle}
\endbibitem

\end{thebibliography}

\end{document}